# Steric Constraints as a Global Regulation of Growing Leaf Shape

*(Kirigami in Leaves)*

Etienne COUTURIER, Sylvain COURRECH du PONT, Stéphane DOUADY

Laboratoire MSC, Matière et Système Complexe
UMR 7057, CNRS & Université Paris-Diderot
10 rue Alice Domon et Léonie Duquet, 75205 Paris Cedex 13, France

**Shape is one of the important characteristics for the structures observed in living organisms. Whereas biologists have proposed models where the shape is controlled on a molecular level [1], physicists, following Turing [2] and d'Arcy Thomson [3], have developed theories where patterns arise spontaneously [4]. Here, we propose a volume constraint that restricts the possible shapes of leaves. Focusing on palmate leaves, the central observation is that developing leaves first grow folded inside a bud, limited by the previous and subsequent leaves. We show that growing folded in this small volume controls globally the leaf development. This induces a direct relationship between the way it was folded and the final unfolded shape of the leaf. These dependencies can be approximated as simple geometrical relationships that we confirm on both folded embryonic and unfolded mature leaves. We find that independently of their position in the phylogenetic tree, these relationships work for folded species, but do not work for non-folded species. This steric constraint is a simple way to impose a global regulation for the leaf growth. Such steric regulation should be more general and considered as a new simple means of global regulation.**

Leaves fascinate for their diversity. They can be simple, with lobes (palmate), with leaflets (compound), or dissected with holes. On one single plant, the shape of leaves vary, sometimes strongly (heterophilly). Despite this diversity, some common features are intuitively guessed. Until now, botanists have proposed two mechanisms to explain leaf forms. The first mechanism is based on localised enhancements and reductions of growth of the free margin of the embryonic leaf [5-6], which create the peaks and the valleys of the leaf border [7-8]. The second mechanism is the death of patches of cells (programmed cell death, PCD) that forms perforations in the leaf during the lamina development. When perforations are positioned near the leaf contour, the marginal tissue eventually breaks, as in *Philodendron Monstrosa* (Araceae, monocotyledon), resulting in a deeply dissected blade (pinnatisect) [9]. A particular case has been described for the dissected shape of palm leaves (Arecaceae, monocotyledons). The leaf first develops with many folds, where PCD eventually takes place, creating cuts [10].

These two mechanisms are general so that they can be tuned to reproduce the final shape of any leaf. They conceptually apply for a flat leaf during its expansion and do not take into account the actual geometry of the growing leaf inside the bud. Only the last case takes into account this geometry, through the folds. Folds have been recently highlighted for their mechanical importance in thin sheets [11-12] and are ascribed to play a role in the expansion of hornbeam leaves [13]. Focusing on palmate leaves, we found that most of them are first growing folded inside a bud.

Leaves are highly organised botanical elements. They originate from small groups of cells (primordia) protruding around the shoot apex (fig. 1a). From the beginning they present a fundamental asymmetry: the side turned toward the stem axis (adaxial) will become the smooth and shiny upper side of the leaf turned toward the light; the other side, turned toward outside (abaxial), present hairs and veins protruding and will become the lower side of the leaf.

Once the leaf has grown, it can be noticed that the final vein pattern is organised, hierarchical, and related to the leaf shape. In palmate leaves, each lobe corresponds to a major vein ending at its tip. Lobes and veins have the same hierarchy: beside the central lobe/vein stand symmetrically lateral lobes/veins that start from the same petiole origin. From these primary veins can originate secondary lobes/veins, and similarly for a rare third order (see figure 3).

In buds, leaves grow in a limited space defined by previous and following leaves. To keep on expanding its future lamina within this confined space, leaves either enroll (convolute, revolute or involute), or fold (plicate and palmate leaves) [14-15].

Our first remark is that in palmate leaves, folds are not irregular but strictly follow the leaf organisation (fig. 1f-g). Following the leaf asymmetry, the folds showing the abaxial epidermis outside (anticlinal folds) are very different than the ones showing the adaxial epidermis outside (synclinal folds). The anticlinal folds coincide with the main veins and follow their hierarchy (fig. 1f). This correlation between the veins and anticlinal folds is strictly inclusive: many veins do not correspond to any fold but an anticlinal fold always corresponds to a main vein. On the contrary, synclinal folds do not correspond to any major vein and are rather crossed only by the smallest ones (fig. 1g). Thus we call them "anti-veins".

Our second and main observation is that the whole perimeter of folded leaves growing inside a bud is located at a particular place. For Palmate leaves it is located at the border of the folding volume, on the adaxial side (fig. 1g). As the leaf margin links the end of synclines to the end of anticlines, this last property induces strong relationships between the shape of the leaf and the way it was folded inside the bud. This relates to Kirigami, where a piece of paper can take any shape by folding it in a particular way, and then cutting it on one side [16].

The first consequence is that in this elongated geometry, with the leaf perimeter turned adaxially, anticlines set peaks (or lobes) and synclines set valleys (or sinuses) of the unfolded leaf. Even more, for the folded leaf, the two sides of the valley superimpose when they join at the end of a fold. The peak and sinus must then be symmetric around the fold. Figure 2 shows this local property on immature leaves extracted from the bud of various species.

These steric constraints not only induce local geometric relationships in immature leaves, but also global ones that are kept in the mature leaves, determining their shape. Using the relation between the folds and the main veins (first and second order), mature leaves can be folded back like we expect they were when growing in the bud. Figure 3 shows various mature leaves folded this way. One can see that the whole leaf perimeter collapses onto a simple curve. The different leaf shapes can then be ascribed essentially to the variation of angles between the veins and to the different curves delimiting the folded leaf.

This property can be translated into quantitative relationships between the length of veins and anti-veins and the angles between them. These relationships, restraining the range of possible shapes, are shown in the case of sycamore leaves in figure 4. For instance, considering two consecutive lobes, the longest is the one whose vein makes the biggest angle with the corresponding anti-vein. Similarly, the smaller of two sinuses has an anti-vein that makes the bigger angle with the intermediate vein. As well as in figure 2 and 3, these relationships hold for palmate species widely separated on the evolutionary scale (APG II classification)[17], while more closely related species with leaves that do not grow folded inside the bud show no trace of such organization (see suppl. Mat.).

While there are many variations of shapes, even on a single plant, each palmate leaf obeys the Kirigami property (the perimeter folding back onto a single curve), and falls along the same constraining geometrical rules. This shows that the leaf shapes, even if varying strongly, are not arbitrary. However, without any particular constraint, differential growth, or apoptosis, can produce any desired shape, even "unnatural". Thus to produce the specific shapes observed requires a particular regulation. This overall limitation of the growth by the constraining volume is the needed regulation.

Even though growing folded in the bud, it is remarkable that the embryonic leaf can be unfolded nearly as flat as the final leaf (fig. 2). This already proves a strong and permanent regulation to keep the leaf lamina flat locally [1]. But to have an overall flat surface, the folds, where this regulation mechanism cannot apply, have also to be considered. If the folds are straight, as elongating veins tend to be, then the overall surface can be flat. A problem remains around the synclinal folds, that are not restricted *a priori* to be straight. Indeed some leaves exhibit curved synclinal folds, and then a characteristic unflatenable place at their sinus.

Based on many observations of dissected buds at different stages (fig. 1), we can write the following scenario for the development of the leaf inside the bud. The primordium first develops, with a central vein and a flat lamina (fig.1a). It extends over the meristem apex, until it reaches the other leaves (fig. 1b). The fact that it stops and will never overpass this limit indicates a contact regulation of the growth [18]. Later on, secondary veins being already there [19], folds appear. We think that these main veins have an active role in the formation of the folds: they rotate the lamina around them, creating the synclinal folds. This pushes the rest of the developing lamina toward the inside, inducing anticlinal folds (fig. 1c). The developing lamina again stops its growth when reaching the volume border, indicating the same contact regulation of the growth (fig. 1d). Further phenomena, like PCD in the case of palm leaves, or inhomogeneous growth in the case of oaks, can also intervene to eventually shape the final leaf form once it is outside the constraining volume of the bud.

The evolutionary interest of such folding mechanisms could be not to regulate the final shape of the leaf, but to protect the very fragile immature leaves. A good way to protect them is to grow them inside buds with protecting scales. In the limited volume of the bud, one way to develop the largest surface, ready to catch light, is to grow folded. We saw that the leaves grow with their own internal dynamics, create folds, and stop when reaching the constraining surface. This overall growth regulation is a simple way to ensure that the whole volume is evenly occupied, and that no space is lost. The asymmetric development of the folds (fig. 1), with the fragile lamina pushed toward the more protected adaxial side, and the more robust veins, often with hairs, covering the whole abaxial side, could also be evolutionary interesting. Not only the bud scales, but also this asymmetrical folding geometry, protect the leaf from cold, dryness, and the numerous small predators (insects). The leaf shape, with its lobes, could just be a secondary consequence of the interplay of the optimization and protective mechanisms, achieved through the folding and the growth contact regulation. This explains the observation of the predominance of palmate leaves in cold-temperate regions [20], where this protection is most needed. The variations in development before outside expansion could finally explain the continuous variation of shapes and the number of lobes.

This global mechanical regulation of the shape could happen in leaves as their shape is, like phyllotaxis [4], not an essential criterion for their reproduction success, contrary to the shape of flowers, that has be more directly controlled [21]. The evolutionary pressure on leaf is to grow as large and as soon as possible, whatever the growth conditions are. It can be more efficiently achieved with the global regulation we propose. The disparity of this Kirigami property along the whole evolutionary tree also shows that is not a highly fixed and stabilized property. This Kirigami theory, deriving from the direct observation of the geometry of immature leaves, reveals many underlying regulations, folding the leaf around the veins, making the leaf margin mechanically sensitive, and insuring the flatness of the leaf. They all deserve more study, and in particular their underlying molecular mechanisms. This shape theory also points out how much the simple physical constraint of growing inside a finite volume can have an important effect, widely underestimated up to now. It finally shows how still little is known of the various shapes and their origin, not to speak of their control.

**References**

1. Nath, U. , Crawford, B. C. W. , Carpenter, R. & Coen, E.  Genetic control of surface curvature. *Science* **299**, 1404 - 1407  (2003)
2.  Turing, A. M.  The chemical basis of morphogenesis. *Phil. Trans. R. Soc.* **237** *B* 37-72 (1952)
3. Thompson, D. A.W. On Growth and Form, 2nd edn (Cambridge Univ. Press, Cambridge, UK, 1942)
4. Douady, S. & Couder , Y. Phyllotaxis as a self organizing process-Part I-II-III. *J. theor. Biol.* **178**  255-312 (1996)
5. Arunika, H. L.,  Gunawardena & A. N. , Dengler, N. G. Alternative modes of leaf dissection in monocotyledons. *Bot. J. Linn. Soc.* **150**, 25-44. (2006)
6. Hagemann, W. & Gleissberg S. Organogenetic capacity of leaves: the significance of marginal blastozones in angiosperms. *Plant. Systemat. Evol.* **199**, 121-152.(1996)
7. Franks, N. R. & Britton, N. F. The possible role of reaction-diffusion in leaf shape. *Proc. R. Soc. Lond. B* **267**, 1295 -1300. (2000)
8.  Marder, M., Sharon, E. , Roman, B., & Smith, S.  Theory of the edge of leaves. *E.P.L.* **62**   498-504 (2003).
9. Melville, R. & Wrigley, F. A. Fenestration in the leaves of Monstera and its bearing on the morphogenesis and colour patterns of leaves. Bot. J. Linn. Soc. **62**: 1-16. (1969)
10. Kaplan, D. R. , Dengler, N. G. & Dengler, R. E. The mechanism of plication inception in palm leaves: problem and developmental morphology.  *Can. J. Bot.* **60** , 2939-2975.(1982)
11. Boudaoud, A., Patrício, P., Couder, Y.  & Ben Amar, M. Dynamics of singularities in a constrained elastic plate. *Nature* **407**, 718-720 (2000).
12. Mahadevan, L. & Rica, S. Self-organized origami, *Science*, **307**, 1740, (2005)
13.  Kobayashi, H. , Kresling, B. & Vincent, J.F.V.  The geometry of unfolding tree leaves. *Proc. R. Soc. Lond. B* **265**, 147-154  (1998)
14.  Clos, D. Monographie de la préfoliation, dans ses rapports avec les divers degrés de classification. (Rouget frères et Delahaut, Toulouse, 1870)
15. Adanson, M. Familles des plantes : contenant une préface historique sur l'état ancien et actuel de la botanique, et une théorie de cette science. ( Chez Vincent imprimeur du Comte de Provence, Paris, 1763)
16. Demaine, E.D. ,  Demaine, M.L. & Lubiw, A. Folding and One Straight Cut Suffice. Proceedings of the 10th Annual ACM-SIAM Symposium on Discrete Algorithms 891-892 (1998)
17. Angiosperm Phylogeny Group . An update of the Angiosperm Phylogeny Group classification for the orders and families of flowering plants: APG II. *Bot. J. Linn. Soc.* **141,** 399-436 (2003)
18. Shraiman, B.I. Mechanical feedback as a mechanism of growth control. *Proc. Natl. Acad. Sci. USA* **102**, 3318-3323 (2005)
19. Scarpella, E. , Marcos, D. , Friml, J. & Berleth, T. Control of leaf vascular patterning by polar auxin transport. *Genes Dev.* **20** , 1015-1027. (2006)

20. Bailey, I.W, Sinott, E.W. The climatic distribution of certain type of angiosperm leaves. *Am. J. Bot.* **3,** 24-39 (1916)
21. Rolland-Lagan, A. G., Bangham, A. J. & Coen, E. Growth dynamics underlying petal shape and assymmetry. *Nature* **422**, 161-163 (2003)

Figures

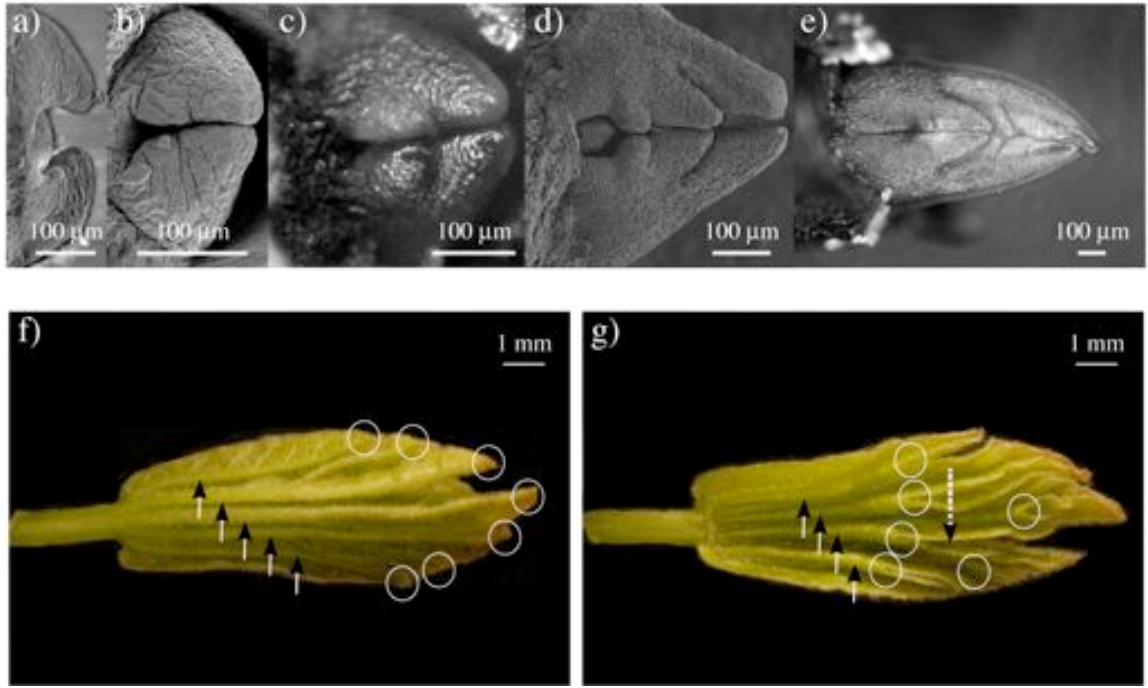

Figure 1

a-e) Developing embryonic leaves of *Acer pseudoplatanus*. For such opposite decussate phillotaxy, symmetric leaves develop at the same time. The primordia first expand over the stem apex (a) and then the two symmetric leaves meet (b), limiting each other in their future growth (c-e). c) A first lateral fold has appeared between the central and the lateral veins. e) One can observe secondary veins. Pictures (a), (b) and (d) are MEB pictures courtesy of Isabel Le Disquet from IFR 83 of UPMC-Paris 6. Pictures (c) and (e) are microscope pictures.

f-g) Folded leaf extracted from a bud of *Acer campestre*. f) The abaxial side shows the anticlinal folds (arrows) running along veins and ending at peaks (circles). g) The adaxial side shows the synclinal folds (arrows) running along (immaterial) anti-veins and ending at sinuses (circles). Peaks and sinuses stand in the contact plane of the pair of leaves (pictures a-e), but while peaks are at the extreme of this contact surface, sinuses are inside.

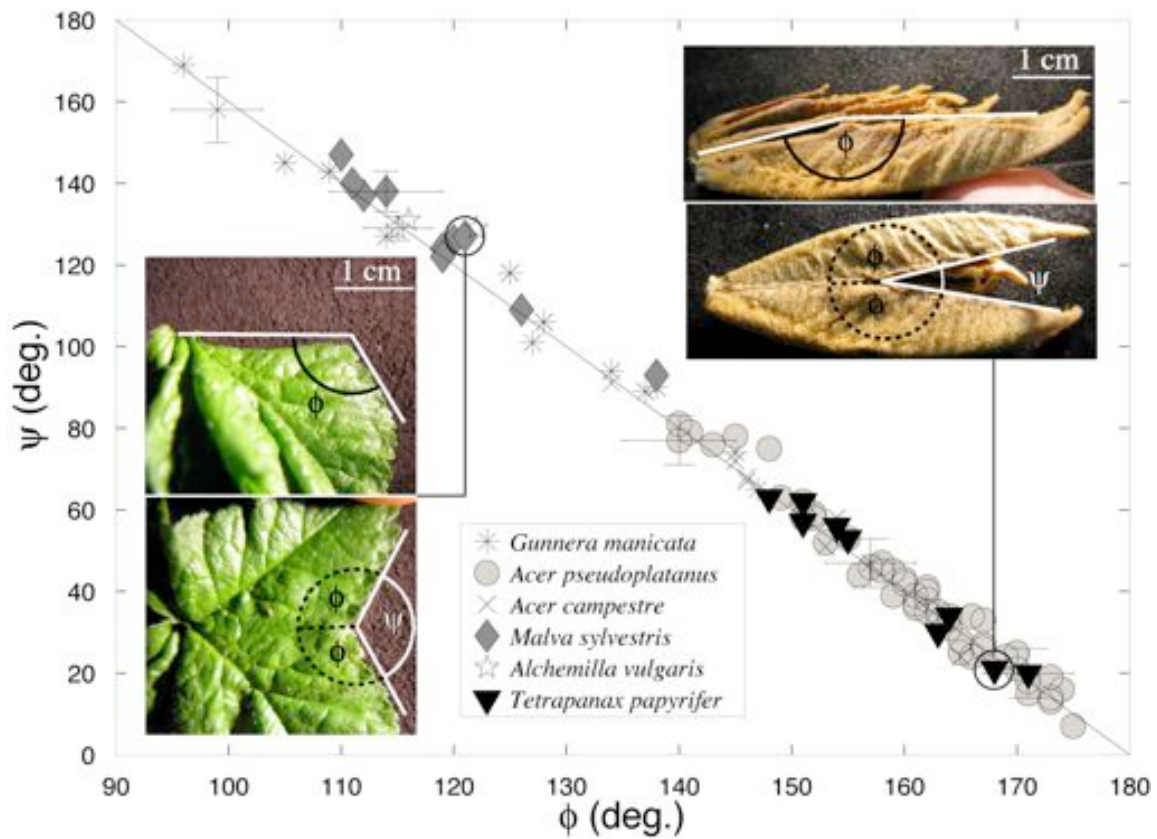

Figure 2

Local relationship between synclinal folds and sinuses on immature leaves before their expansion outside the bud. Synclinal fold angles $\phi$ are measured on folded leaves extracted from buds. Leaves are then unfolded to measure the corresponding sinus opening angle $\psi$ as illustrated by the insets. The fold, setting the sinus whose contours locally superimpose, is the symmetry axis of the sinus making $\psi$ to be: $\psi = 360° − 2\phi$ (line). The studied palmate leaf species belong to different order of eudicots. Following the APG II classification *Acer pseudoplatanus* and *Acer campestre* (Sapindales), *Malva sylvestris* and *Alchemilla vulgaris* (Malvales) are Rosids, *Tetrapanax papyriferum* (Apiales) is an Asterid and *Gunnera manicata* is a Gunnerale.

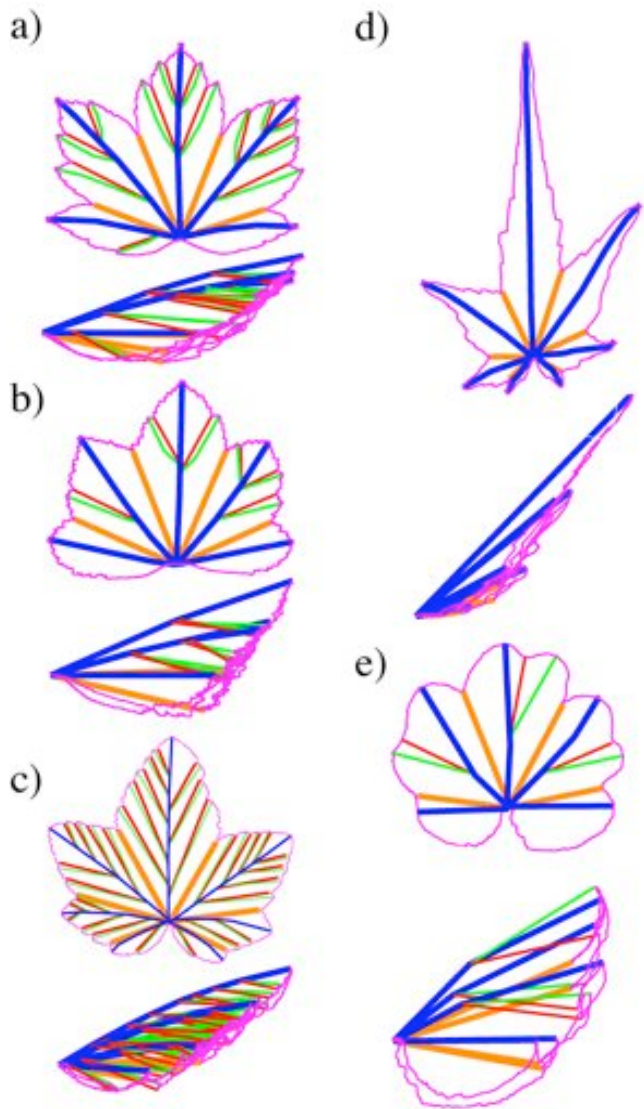

Figure 3

Mature leaves of different species, numerically folded back.
The leaf contour is pink. Primary and secondary veins are respectively blue and green. Primary and secondary anti-veins are yellow and red respectively. Only veins ending at peaks are represented and stand for anticlinal folds that take place along segments linking two consecutive branching points or a branching point to a peak. Synclinal folds run along segments (anti-veins) linking a sinus to the branching point of the two surrounding veins. The thickness of the leaf is not taken into account and the leaf is folded back onto a plane, holding the angles to the best (see supplementary material).
a) *Acer pseudoplatanus*, b) *Malva sylvestris, manicata*, c) *Ribes nigrum*, d) *Sida hermaphrodita*, e) *Gunnera*. Following the APG II classification , these species belong to different orders of core eudicots: *Acer pseudoplatanus* (Sapindales), *Sida hermaphrodita* and *Malva sylvestris* (Malvales) are Rosids, *Gunnera manicata* is a Gunnerale and *Ribes nigrum* is a Saxifragale.

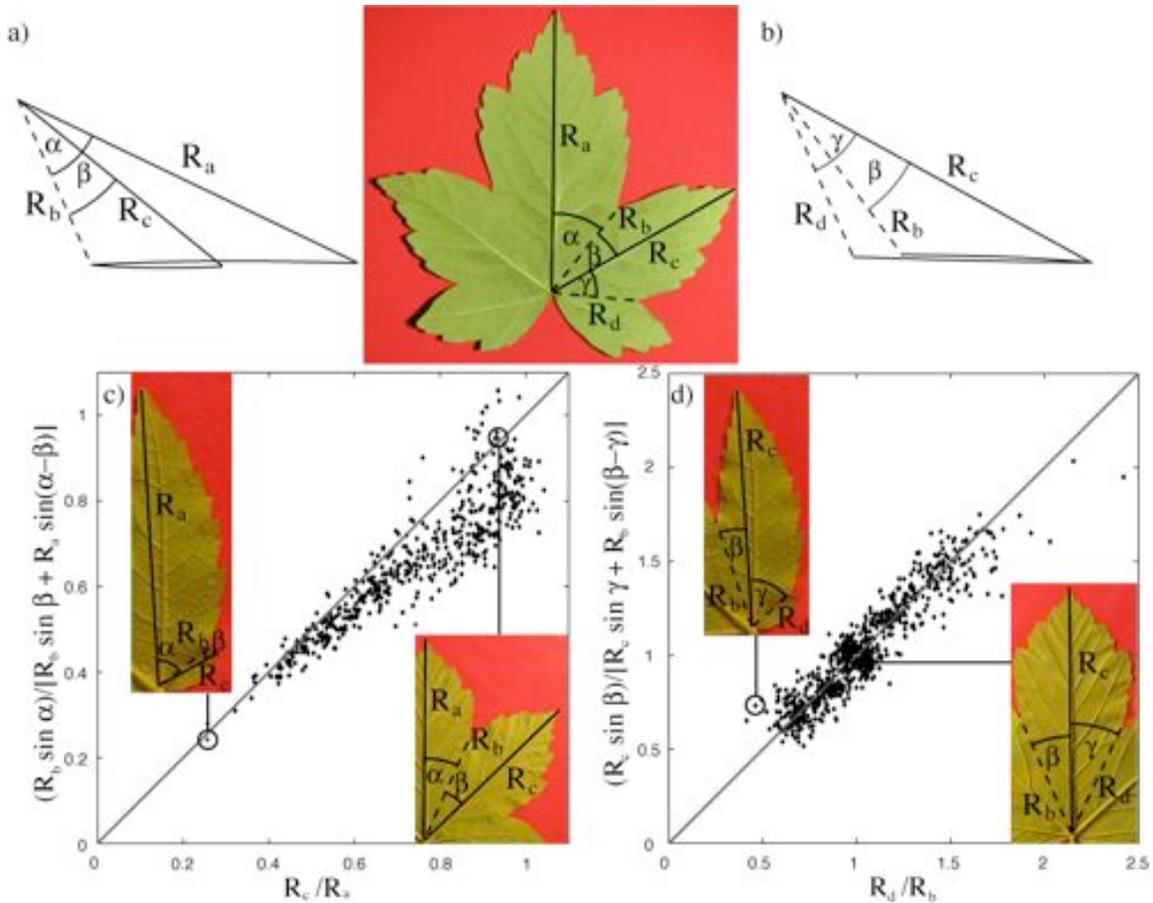

Figure 4

Geometric relationships between two successive lobes and sinuses coming from the Kirigami property.

a) Two consecutive primary lobes have veins of lengths $R_a$ and $R_c$. They are respectively making an angle $\alpha$ and $\beta$ with the anti-vein, of length $R_b$, between them. b) The vein of length $R_c$ is surrounded by two anti-veins of lengths $R_b$ and $R_d$. These are respectively making an angle $\beta$ and $\gamma$ with the vein. According to figure (a) where the folded "leaf" has its contour onto a line, the ratio $R_c / R_a$ of lengths of two consecutive lobes writes in function of all parameters but $R_c$: $R_c / R_a = (R_b \sin \alpha) / [R_b \sin \beta + R_a \sin (\alpha - \beta)]$. The ratio $R_d / R_b$ of lengths of two consecutive anti-veins, as illustrated by figure (b), writes in function of all parameters but $R_d$: $R_d / R_b = (R_c \sin \beta) / [R_c \sin \gamma + R_b \sin (\beta - \gamma)]$ (formula derivation are given in supplementary material). These predictions of the Kirigami property are compared to the measured values on figure (c) for veins and on figure (d) for anti-veins. Points represent 121 sycamore leaves. For each leaf, all pair of consecutive primary veins or anti-veins (four for the leaf presented) have been measured.

**Supplementary materials**

Derivation of formula of Figure 4.
Taking the sketch of figure 4 (a), the condition that contour superimpose is:
$$\left(\vec{R}_a - \vec{R}_c\right) \wedge \left(\vec{R}_a - \vec{R}_b\right) = \vec{0}$$
which writes
$$-\vec{R}_a \wedge \vec{R}_b - \vec{R}_c \wedge \vec{R}_a + \vec{R}_b \wedge \vec{R}_c = \vec{0}$$
or
$$R_a R_c \sin(\alpha - \beta) + R_c R_b \sin(\beta) = R_a R_b \sin(\alpha).$$
This gives
$$R_c = \frac{\left(R_a R_b \sin(\alpha)\right)}{\left(R_a \sin(\alpha - \beta) + R_b \sin(\beta)\right)}$$
or similarly
$$\frac{R_c}{R_a} = \frac{\left(R_b \sin(\alpha)\right)}{\left(R_a \sin(\alpha - \beta) + R_b \sin(\beta)\right)}.$$
Formula of Figure 4 (d) can be derived in the same way.

Supplementary figures

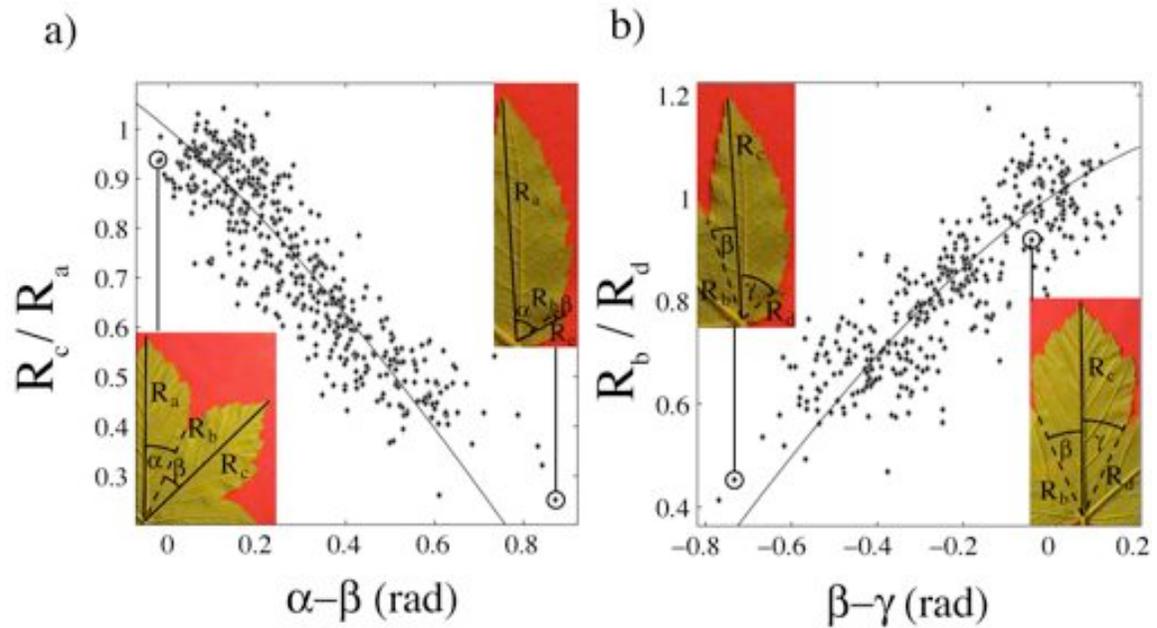

Supplementary figure 1

Quantitative relationship on two successive lobes and sinuses coming from the Kirigami property, for 121 *Acer pseudoplatanus* (sycamore) leaves.
a) Length ratio ($R_a / R_c$) of two consecutive main veins in function of the difference ($\alpha - \beta$) between the angles there are making with the anti-vein. The anti-vein always runs closer to the smaller lobe (right hand-side picture). When lobes have equal length the anti-vein is then a bisecting line ($\alpha = \beta$, left hand-side picture).
b) Length ratio ($R_b / R_d$) of two consecutive main anti-veins in function of the difference ($\beta - \gamma$) between the angles there are making with the vein. The vein always runs closer to the smaller anti-vein (right hand-side picture). When anti-veins have equal length the vein is then a bisecting line ($\beta = \gamma$, left hand-side picture).

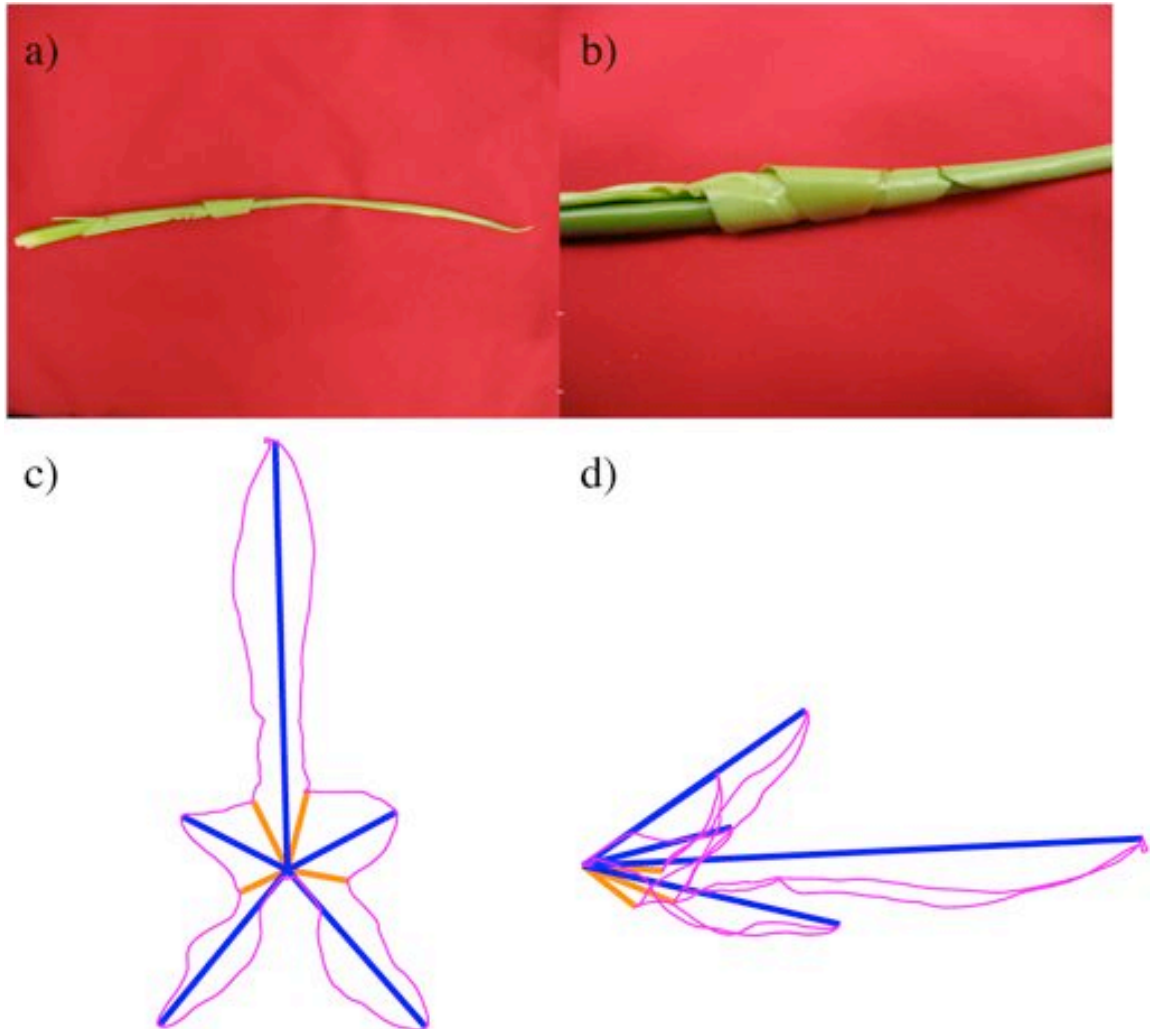

Supplementary figure 2

Example of an enrolled leaf: the *Philodendron bipenifolium* (Araceae).
a) A juvenil *Philodendron bipenifolium* leaf. b) Detail of the enrolled leaf. The leaf does not grow folded but enrolled ; it is an involute leaf.
c) A mature *Philodendron bipenifolium* leaf. d) The same leaf numerically folded. Like for folding of figure 4, the thickness of the leaf is not taken into account and the leaf is folded onto a plane, holding the angles to the best (see supplementary figure 5 for a detailed explanation of the method). As the leaf does not grow folded, it does not obey the Kirigami property and is not foldable with its contour lying on a simple curve.

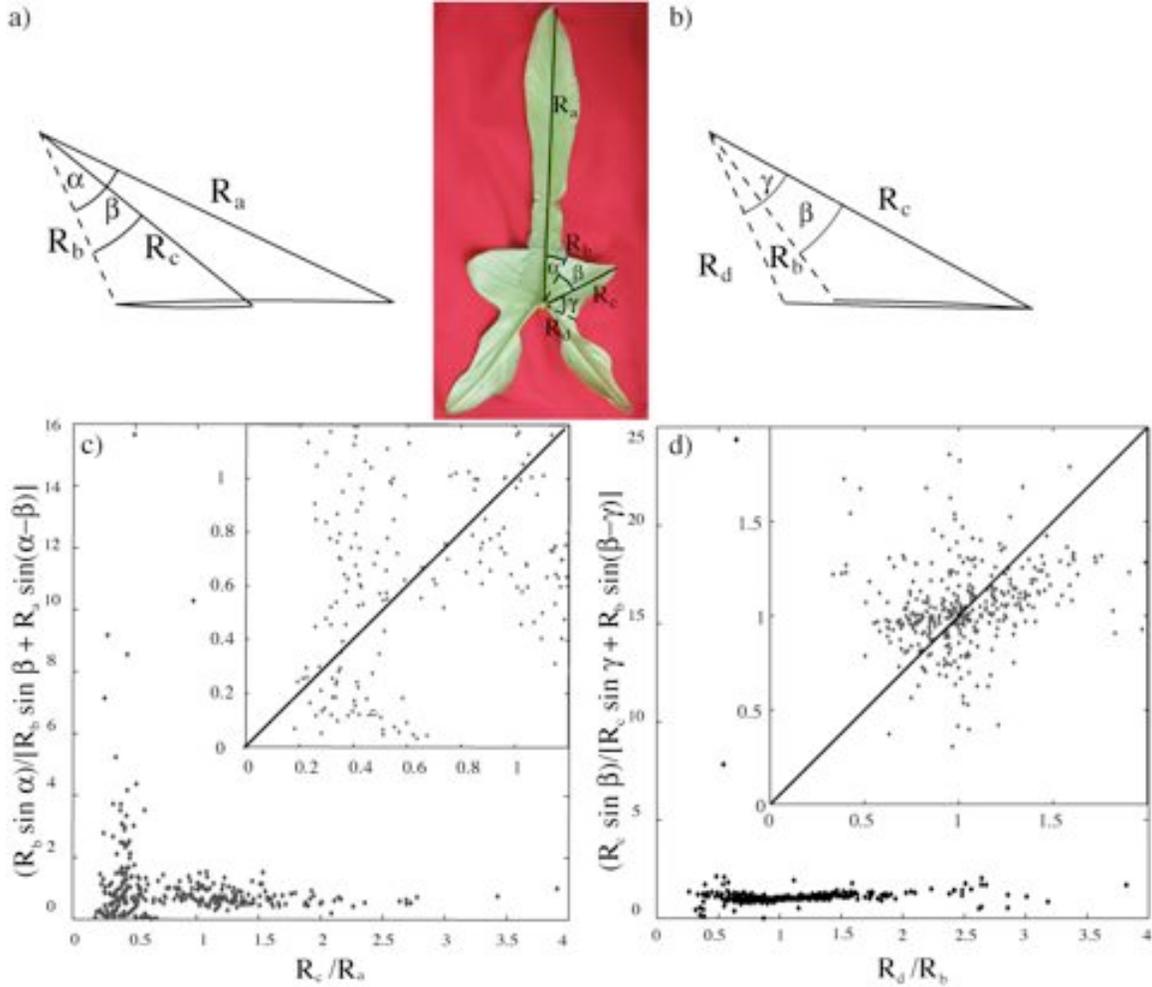

Supplementary figure 3

Geometric relationships between two successive lobes and sinuses coming from the Kirigami property, for the enrolled *Philodendron bipenifolium* leaf.

a) Two consecutive primary lobes have veins of lengths $R_a$ and $R_c$. They are respectively making an angle $\alpha$ and $\beta$ with the anti-vein, of length $R_b$, between them. b) The vein of length $R_c$ is surrounded by two anti-veins of lengths $R_b$ and $R_d$. These are respectively making angle an $\beta$ and $\gamma$ with the vein. According to figure (a), in the frame of the Kirigami property, the ratio $R_c / R_a$ of lengths of two consecutive lobes writes in function of all parameters but $R_c$: $R_c / R_a = (R_b \sin \alpha) / [R_b \sin \beta + R_a \sin (\alpha - \beta)]$. The ratio $R_d / R_b$ of lengths of two consecutive anti-veins, as illustrated by figure (b), writes in function of all parameters but $R_d$: $R_d / R_b = (R_c \sin \beta) / [R_c \sin \gamma + R_b \sin (\beta - \gamma)]$.

These predictions of the Kirigami property are compared to the measured values on figure (c) for veins and on figure (d) for anti-veins. Points represent 85 *Philodendron bipenifolium* leaves. For each leaf, all pair of consecutive primary veins or anti-veins (four for the presented leaf) have been measured. One observes that points are scattered in both figures. This figure, as supplementary figure 2, shows that the Kirigami property is shape demanding.

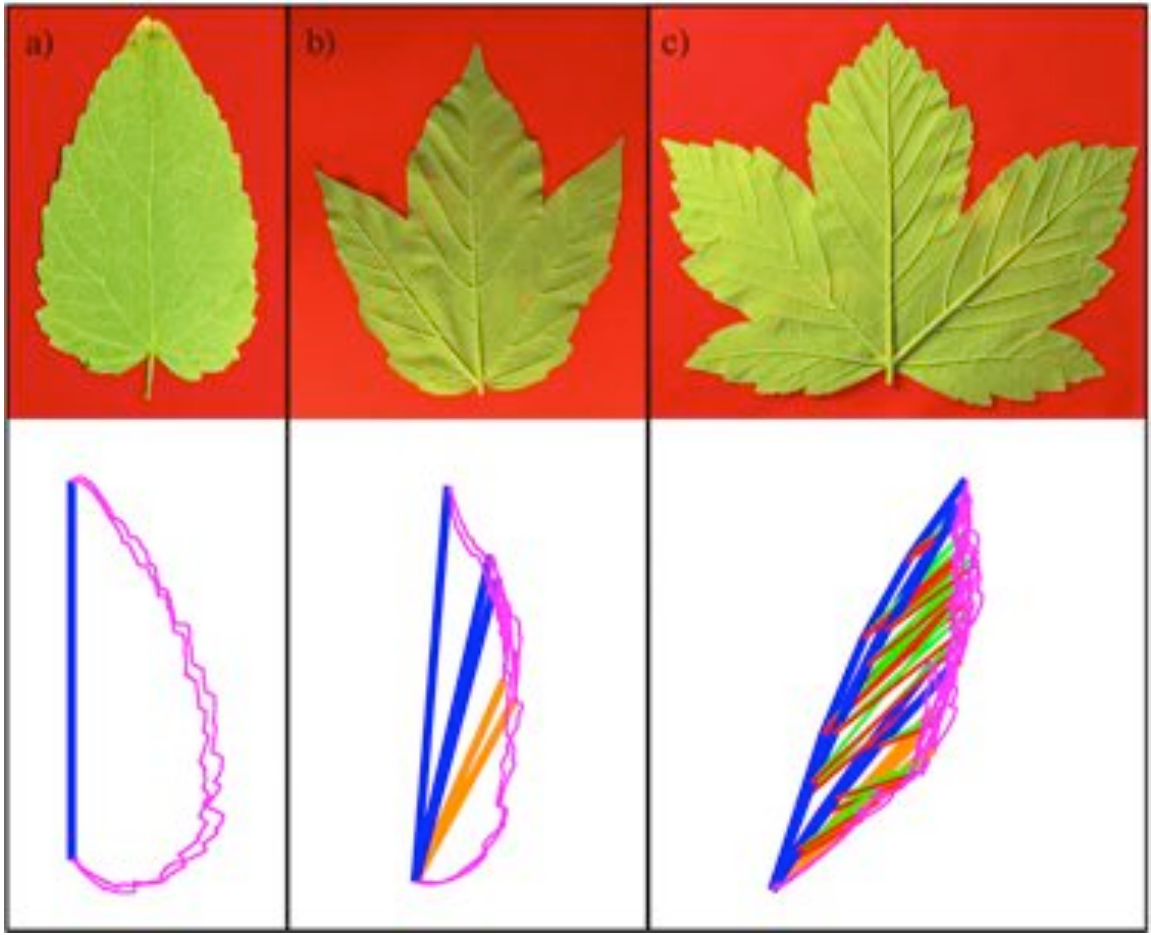

Supplementary figure 4

*Acer pseudoplatanus* leaves: pictures and numerical folding.
a) One lobe leaf. b) Three lobe leaf. c) Five lobe leaf.

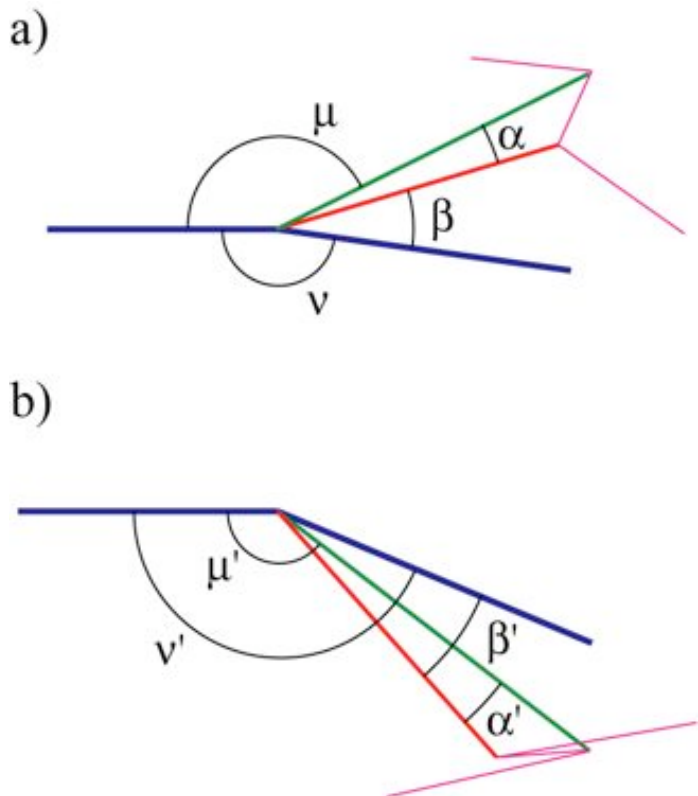

a)

b)

Supplementary figure 5

Numerical folding method.
Branching detail: unfolded (a) and folded in a plane (b). Blue and green segments are anticlinal folds while the synclinal fold is red. The "leaf" contour is pink.
As seen on figure (a), the sum of all angles is $\alpha + \beta + \mu + \nu = 2\pi$. Considering the angle $\nu$', the figure (b) is in a plane only if: $\nu' = \mu' + \beta' - \alpha'$. If the sum of angles is equal to $2\pi$, this equation rewrites: $\nu' = \pi - \alpha'$. Folding figure (a) in figure (b) with keeping to the best the angle $\nu$ is minimizing the quantity $(\nu' - \nu)^2 + (\alpha' - \alpha)^2$ which rewrites $(\nu' - \nu)^2 + (\pi - \nu' - \alpha)^2$. One finds the best $\nu'$ value: $\nu' = (\pi + \nu - \alpha)/2$. In the same way, one find: $\alpha' = (\pi + \alpha - \nu)/2$, $\beta' = (\pi + \beta - \mu)/2$ and $\mu' = (\pi + \mu - \beta)/2$.

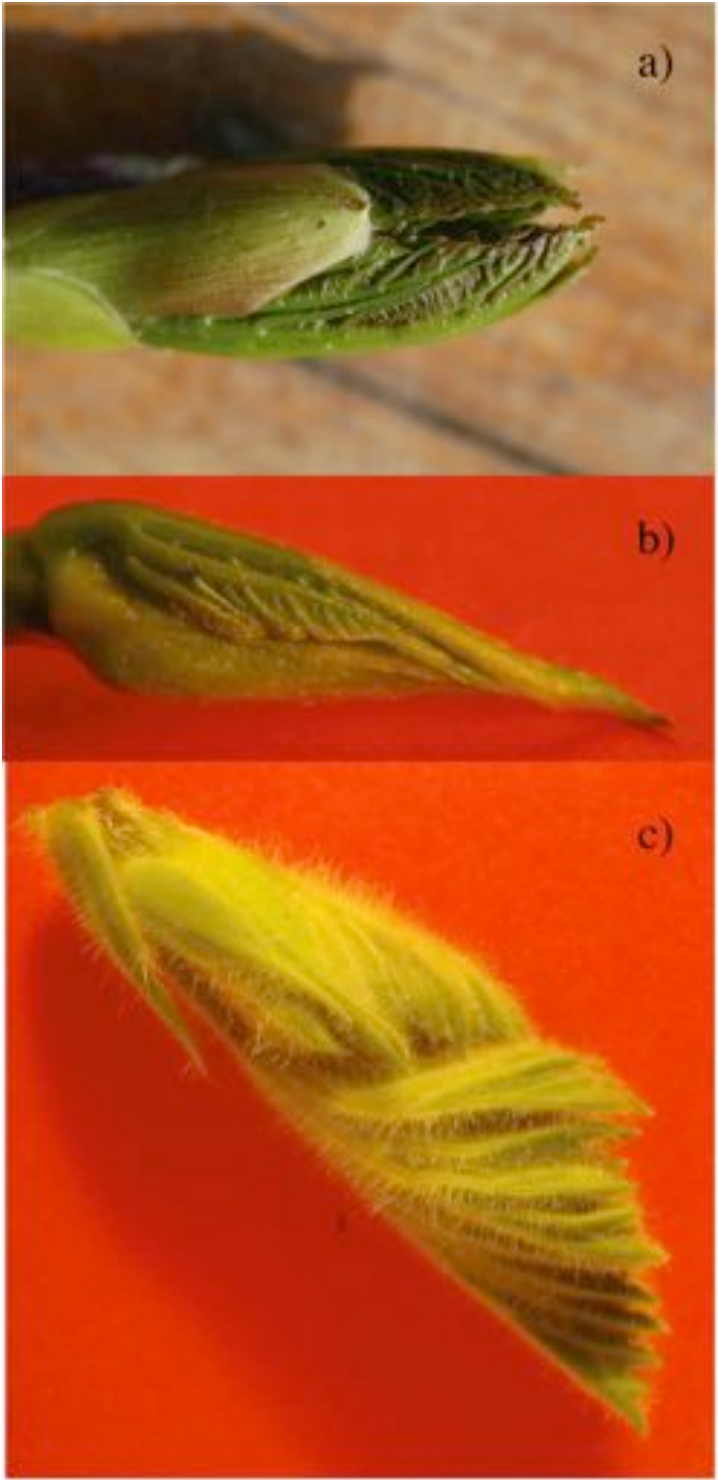
Supplementary figure 6

Dissected Kirigami buds showing the packed folded leaves. a) *Acer pseudoplatanus* (Sapindales), b) *Murus platanifolium* (Rosales), c) *Pelargonium cuculatum* (Geraniales)

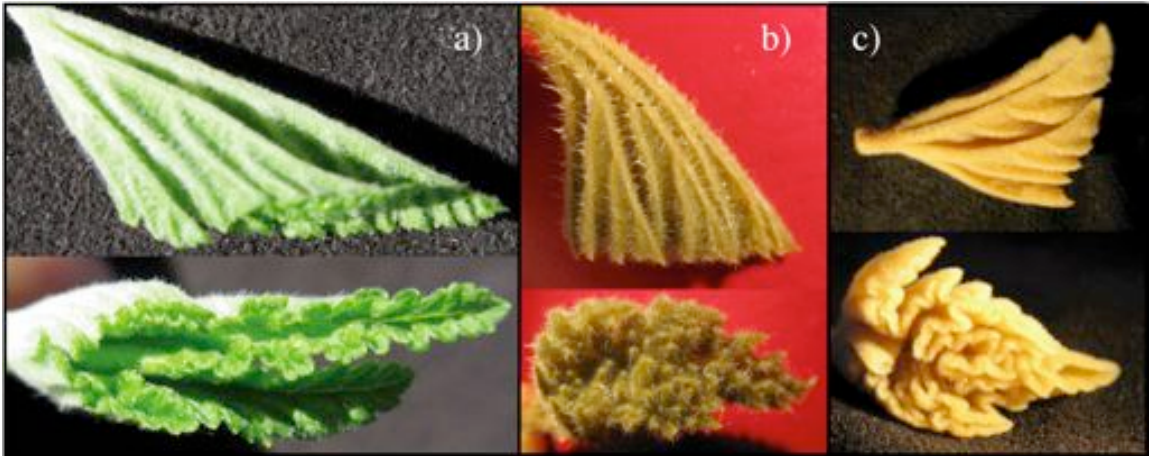

Supplementary figure 7

Side and front views of Kirigami leaves. The front view shows the adaxial contact plane.
a) *Ribes nigrum* (Saxifragales) b) *Pelargonium cuculatum* (Geraniales) c) *Malva sylvestris* (Malvales).